\documentclass[11pt]{article}
\usepackage[T1]{fontenc}
\usepackage[utf8]{inputenc}
\usepackage{lmodern}
\usepackage[letterpaper,margin=1in]{geometry}
\usepackage{amsmath,amssymb,booktabs,graphicx,array,longtable}
\usepackage{microtype}
\usepackage[hidelinks]{hyperref}
\hypersetup{
  pdftitle={Abstention Errors and Segment Support in CUAD: An Auditable Contract-Extraction Case Study},
  pdfauthor={Zeki Emre Tekin},
  pdfsubject={Contract-extraction evaluation on CUAD},
  pdfkeywords={CUAD, contract extraction, abstention, evaluation, reproducibility}
}
\usepackage{caption}
\newcommand{\MainPrecision}{8.2\%}
\newcommand{\MainRecall}{82.2\%}

\newcommand{\CandidateMatch}{20.3\%}
\newcommand{\NoAnswerRate}{45.4\%}
\newcommand{\NoAnswerShare}{64.2\%}
\newcommand{\NoAnswerFPShare}{80.6\%}
\newcommand{\MainTP}{2,172}
\newcommand{\MainFP}{24,272}
\newcommand{\MainFN}{471}
\newcommand{\MainReturned}{30,464}
\newcommand{\MainMatched}{6,192}

\title{Abstention Errors and Segment Support in CUAD:\\An Auditable Contract-Extraction Case Study}
\author{Zeki Emre Tekin\\\small Independent researcher}
\date{7 September 2026}
\begin{document}
\maketitle

\begin{abstract}
Contract-clause extraction benchmarks measure reference recovery, but interpreting a system for review assistance also requires measuring unnecessary output and the evidence available within categories. This paper presents a retrospective, reproducible evaluation of a fixed public RoBERTa checkpoint on CUAD's published 102-contract test split. Candidate generation is preserved from an earlier frozen run; matching errors are corrected and category thresholds are reselected on 62 training-split contracts using the original 90\% recall target. The corrected operating point recovers \MainRecall{} of reference spans at \MainPrecision{} CUAD precision. Of \MainReturned{} returned candidate strings, 19,562 occur on questions with no annotated answer. The system answers 1,335 of 2,938 such questions, a \NoAnswerRate{} false-positive rate (95\% contract-bootstrap interval: 43.3--47.9\%). Matching candidates constitute \CandidateMatch{} of returned strings, illustrating why reference-based precision and candidate-level review burden require different denominators. Only nine of 41 categories have at least 30 positive and 30 negative contracts; this count changes to 18 and two under support minima of 20 and 40. Two categories have no negative contracts, making their no-answer false-positive rates undefined. These findings describe one checkpoint, decoder, and threshold policy. Earlier test access and uncertain checkpoint training overlap limit confirmatory interpretation. The paper supplies the original and corrected analyses, predictions, and software checks as ancillary artifacts; it establishes neither a new release criterion nor production readiness.
\end{abstract}

\section{Introduction}
An extractive contract-review assistant returns passages for a person to inspect. Its usefulness depends on how many relevant passages it recovers, how much unnecessary text it returns, and which types of contracts and clauses support reliable measurement. These questions involve different units: a reference passage, a returned candidate string, a clause question, and a contract. An aggregate metric can be internally consistent while answering only one of them.

The Contract Understanding Atticus Dataset (CUAD) provides an expert-annotated benchmark for this setting \cite{cuad}. Its original evaluation includes category-level results and demonstrates substantial variation between clause types. The present study extends that descriptive perspective by examining no-answer behavior, distinguishing candidate workload from reference recovery, and reporting the positive and negative contract support available for each category. The setting is extraction for review assistance, rather than legal advice or autonomous contract assessment.

The contribution is a reproducible empirical case study of one public checkpoint under a documented operating policy. It has three components: an output decomposition that isolates answering on no-answer questions; a metric-specific support inventory with an explicit sensitivity analysis; and a correction record showing how matching semantics change reported performance. The positive/negative support distinction and the abstention problem are established evaluation concepts. This paper does not propose a new statistical estimator or a generally validated release-readiness method.

The study incorporates a retrospective correction. A protocol and implementation addendum preceded the recorded final inference pass, but the surviving project documents disclose earlier test access, and the original scorer was not identical to the published CUAD implementation. We preserve the original artifacts and clearly separate their results from the corrected analysis. This boundary is necessary for interpreting the evidence, rather than a claim of prospective confirmation.

\section{Related work}
CUAD frames contract review as extraction of passages responsive to a clause category, allowing multiple references and unanswerable questions \cite{cuad}. Its published category analysis is the immediate baseline for the present work's scope. Here, the added emphasis is on output counts on negative questions and the support available for those measurements, not the discovery that category performance varies.

Unanswerable-question evaluation is well established in reading comprehension. SQuAD~2.0 was designed to test systems that might otherwise select plausible text when a question has no answer \cite{squad2}. Our no-answer subset is defined by CUAD's annotations; its error rates should not be compared numerically with those of a differently constructed reading-comprehension benchmark. Selective classification studies the tradeoff between prediction coverage and error risk \cite{selective}. We measure one form of erroneous answering but neither implement that work's risk-control procedure nor claim its guarantees.

Structured evaluation methods such as CheckList organize tests around capabilities and behaviors that aggregate accuracy can miss \cite{checklist}. Our segments are observational slices of an existing benchmark, not newly generated behavioral tests. Dataset documentation work emphasizes communicating collection conditions, intended uses, and limitations \cite{datasheets}. These concerns motivate reporting the exact CUAD files used, the expanded structure of the threshold-selection data, and the distinction between prior model training and selection performed for this study. The NIST AI RMF likewise motivates context-sensitive measurement and documentation \cite{nist}; it supplies neither this study's operating threshold nor its support minimum.

\section{Study design and correction boundary}
\label{sec:design}
\subsection{Recorded history}
Protocol ZET-EVAL-P-001 is dated 5 August 2026. These study-local identifiers use the author's initials, ZET; they are document-control labels, not external registration numbers. An implementation addendum dated 20 August specifies the checkpoint, candidate handling, threshold fallback, segments, and metric-specific support rules. The final original run manifest records a freeze at 06:58:28 UTC on 21 August; its prediction seal records 09:45:42 UTC that day. The original code, manifests, predictions, and outputs are retained as run ZET-EVAL-R-001.

The project README and post-execution audit disclose access to the test file through an earlier calibration path. The earlier audit describes a script targeting the test file and no dataset or predictions present at its review, but the available record does not fully reconstruct the earlier access or what was examined. Separately, the addendum expressly permits test annotations to be read for segment membership and support counts before prediction. Consequently, this paper makes no claim that the test set was never accessed or inspected. The verifiable boundary is consistency of the preserved final files with their recorded hashes; local timestamps and hashes do not establish externally witnessed preregistration.

The present analysis, ZET-EVAL-A-002, corrects matching and reporting after inspecting the original results. Its correction plan was written before implementing this reanalysis, with the already known discrepancies disclosed. It is retrospective, and it does not restore an independent holdout. Appendix~\ref{app:correction} reports the original operating point, a scoring-only correction at the original thresholds, and the corrected operating point used in the main text.

\subsection{Data and units}
CUAD v1 contains 510 contracts and 41 clause categories \cite{cuad}. We preserve the published split of 408 training contracts and 102 test contracts. A fixed seed, 20260805, selects 62 of the training contracts, stratified by a coarse family inferred from document titles, for threshold selection. The remaining 346 contracts are not used to train a new model in this study.

The selection input is the published \texttt{train\_separate\_questions.json}. It expands positive annotations into separate question rows. The 62 selected contracts therefore contain \textbf{3,386 rows}, representing \textbf{2,542 distinct contract/category pairs}, and 1,634 reference spans. These rows must not be described as independent questions. Repeated positive questions can reuse the same context and candidate output while carrying different reference annotations. We retain this source representation for the original threshold rule; selection-set precision is not presented as a performance estimate.

The test input is the published \texttt{test.json}, with 4,182 questions, one for each of 41 categories in each of 102 contracts. There are 2,643 reference spans on 1,244 positive questions and 2,938 questions with no annotated answer. Each test question uses the references in that exact file, not those substituted from another CUAD serialization. All saved reference lists and candidate character offsets were checked against their recorded source files. Document and reference lengths in the segment registry use Python whitespace-separated tokens, not model subword tokens.

\subsection{Checkpoint and candidate generation}
The system is \texttt{akdeniz27/roberta-base-cuad}, pinned at revision
\begin{center}\small\texttt{217bb9b5c398658157a0d906df3577643507671e}.\end{center}
Its model card identifies CUAD fine-tuning and links the original checkpoint source, but does not establish the exact contract-level training membership \cite{modelcard}. Since our selection contracts come from CUAD's training split, their independence from checkpoint training is unverified and training overlap is plausible. We therefore call them the \emph{threshold-selection set}, not an independently held-out validation set. The model card also does not establish complete test-exposure history. The results describe performance on the published test inputs, not a proven absence of prior exposure.

The preserved decoder uses maximum sequence length 512, overlap stride 128, maximum answer length 512 tokens, and 20 candidate start and end positions per window. Valid start/end combinations are ranked by summed logits; repeated character locations retain their maximum score; the top 20 locations across windows are saved. Distinct text strings are returned after thresholding, so multiple overlapping variants can remain while identical strings are counted once per question.

For candidate $j$ of question $q$, the score is
\begin{equation}
 c_{qj}=s_{qj}+e_{qj}-\min_{w\in W_q}\bigl(s_{qw,0}+e_{qw,0}\bigr),
\end{equation}
where $s$ and $e$ are start and end logits and position 0 represents the null answer. Each candidate has its own score. This uncalibrated logit difference is not a probability. The recorded processing totals are 165,951 selection-set and 140,359 test feature windows, distinct from the number of batched model calls. The reanalysis uses the saved candidates without new inference.

\subsection{Matching and threshold selection}
\label{sec:matching}
The corrected matcher follows the public CUAD evaluator at the revision identified in the ancillary source record \cite{evaluator}. It removes periods, commas, semicolons, and colons, lowercases text, replaces slashes with spaces, and splits on literal spaces to form word sets. Articles are retained. A candidate matches when its word-set Jaccard similarity to a reference is at least 0.50. For the \emph{Parties} category, the official implementation also accepts a raw reference string contained in the prediction. We retain this category-specific rule for compatibility; it is not a legal-semantic validity judgment.

For each category, candidate scores on the original selection set are scanned in descending order. We choose the highest cutoff achieving 90\% reference recall under corrected matching. If the target is unattainable, the lowest available cutoff is retained and the shortfall recorded. A category with no selection references receives no threshold and returns no test candidates. Corrected selection yields 28 categories attaining the target, 12 using the minimum-cutoff fallback, and one with no references (\emph{Source Code Escrow}). The only numeric threshold changed from the original record is \emph{Document Name}, from 8.724527 to 7.479274.

This policy seeks high empirical recall; it does not constrain the false-positive rate or certify population recall. For categories unable to reach the target, its minimum-cutoff fallback can produce substantial output. No alternative fallback, decoder, or target was optimized on the test results. The corrected thresholds were selected after test results were known, using selection data only and the unchanged rule; their provenance is explicitly retrospective.

\section{Measures, uncertainty, and support}
\subsection{Reference, candidate, and question measures}
Let $G_q$ be the reference spans and $P_q$ the distinct returned candidate strings for question $q$. Following CUAD counting, $\mathrm{TP}$ is the number of references matched by at least one candidate, $\mathrm{FN}$ counts unmatched references, and $\mathrm{FP}$ counts candidates matching no reference. Thus,
\begin{equation}
 P_{\mathrm{CUAD}}=\frac{\mathrm{TP}}{\mathrm{TP}+\mathrm{FP}},\qquad
 R=\frac{\mathrm{TP}}{\mathrm{TP}+\mathrm{FN}},\qquad
 F_1=\frac{2\mathrm{TP}}{2\mathrm{TP}+\mathrm{FP}+\mathrm{FN}}.
\end{equation}
One candidate may recover multiple references, and several candidates may match one reference. Consequently, $\mathrm{TP}+\mathrm{FP}$ need not equal the number of returned strings. We separately report the \emph{candidate match fraction}: the number of returned strings matching any reference divided by all returned strings. Candidate ratios are proxies for review burden; no user interface, deduplication workflow, or review time was evaluated.

On the no-answer subset, the question-level false-positive rate is
\begin{equation}
 \mathrm{FPR}_{\varnothing}=
 \frac{\sum_q\mathbf{1}\{|G_q|=0,\ |P_q|>0\}}
 {\sum_q\mathbf{1}\{|G_q|=0\}}.
\end{equation}
This is different from the fraction of returned strings arising on negative questions and from the fraction of unmatched candidates arising there. It is undefined if a category has no negative questions. An unmatched candidate can also occur on a positive question; the absence of negative contracts does not make all span-level false positives unobservable.

\subsection{Uncertainty}
For aggregate and segment ratios we resample contracts with replacement within the relevant segment, retaining all their question counts. We use 1,000 percentile bootstrap draws with seed 20260805 and report the 2.5th and 97.5th percentiles. The machine-readable report includes the number of usable draws when a denominator becomes zero. Thresholds and predictions are fixed within a resample, so these intervals omit uncertainty from checkpoint fitting, threshold selection, annotation, and choice of analysis. They are descriptive marginal intervals, not simultaneous guarantees across the registered slices.

For a single category, each negative contract contributes one binary no-answer outcome. The supplement also provides Wilson intervals for that conditional rate \cite{wilson}. These are useful when all observed outcomes are identical: a zero-error percentile bootstrap can be degenerate and should not imply zero population risk. We do not interpret the mixed reference/candidate precision denominator as independent Bernoulli trials.

\subsection{Prespecified slices and minimum-support screen}
The original registry contains 541 rows across category (S1), clause-family (S2), raw title-derived type (S3), document length (S4), training frequency (S5), answer status (S6), longest reference length (S7), and selected intersections (S8), plus eight supplemental normalized-type rows (S3N). We retain their exact memberships and report every row in the supplement.

Two implementation details constrain interpretation. First, S3 is a string parsed from a title, not an authoritative contract-type annotation; its fragmentation partly reflects that parser. S2 likewise uses the original implementation's fixed category grouping, retained for audit rather than asserted to reproduce an authoritative taxonomy. Second, S5 counts \emph{positive expanded training rows}, not distinct positive contracts. Its frozen cut points are 104 and 332: rare at most 104, medium 105--332, and common more than 332. These labels denote training annotation frequency. They are not a test of contract-level prevalence independent of annotation multiplicity.

The original addendum sets minimum support at 30 positive contracts for recall and 30 negative contracts for no-answer analysis. Both counts must reach 30 for the category to pass both arms. This is a prespecified support screen, not a derivation of the sample size needed for a specified error tolerance. For illustration, 27 successes in 30 independent binary trials has a 90\% point estimate but a 95\% Wilson interval of approximately 74.4--96.5\% \cite{wilson}. Passing a count threshold therefore does not establish 90\% population recall. We add a retrospective sensitivity check at minima of 20 and 40, without replacing the original 30-contract screen.

\section{Results}
\subsection{Corrected operating point}
The corrected operating point recovers \MainTP{} of 2,643 references and returns \MainReturned{} candidate strings. There are \MainFP{} unmatched candidates and \MainFN{} missed references. Table~\ref{tab:metrics} reports reference-based metrics and the separate candidate and no-answer measures. These operating-point metrics use the 90\%-recall target and its fallback rules; they are not directly comparable with CUAD's published AUPR baseline, which summarizes a precision--recall curve across cutoffs (Appendix~\ref{app:other}).

\begin{table}[htbp]\centering
\caption{Corrected test operating point. Intervals resample contracts, conditional on the fixed checkpoint, candidates, and corrected thresholds.}\label{tab:metrics}
\begin{tabular}{lrr}
\toprule
Measure & Estimate & 95\% interval \\
\midrule
CUAD precision & 8.2\% & 7.4--9.1\% \\
Reference recall & 82.2\% & 78.8--85.5\% \\
CUAD F1 & 14.9\% & 13.5--16.3\% \\
Candidate match fraction & 20.3\% & 18.6--22.0\% \\
No-answer question FPR & 45.4\% & 43.3--47.9\% \\
\bottomrule
\end{tabular}

\end{table}

There are \MainMatched{} matching candidate strings, giving a \CandidateMatch{} candidate match fraction compared with \MainPrecision{} CUAD precision. The same queue contains 3.92 unmatched candidates per matching candidate and 11.17 unmatched candidates per recovered reference. It returns 7.28 candidates per question, or 14.03 per recovered reference. These denominators answer different operational questions; none is a measured time saving.

\emph{Source Code Escrow} illustrates a different outcome: the policy returns no candidates on any of its 102 test questions because it has no selection references. Its sole positive test contract contains five reference spans, all missed, giving 0\% recall and undefined precision. Its 0\% no-answer FPR therefore accompanies complete abstention, not successful extraction.

\subsection{Answering on no-answer questions}
\begin{figure}[htbp]\centering
\includegraphics[width=\linewidth]{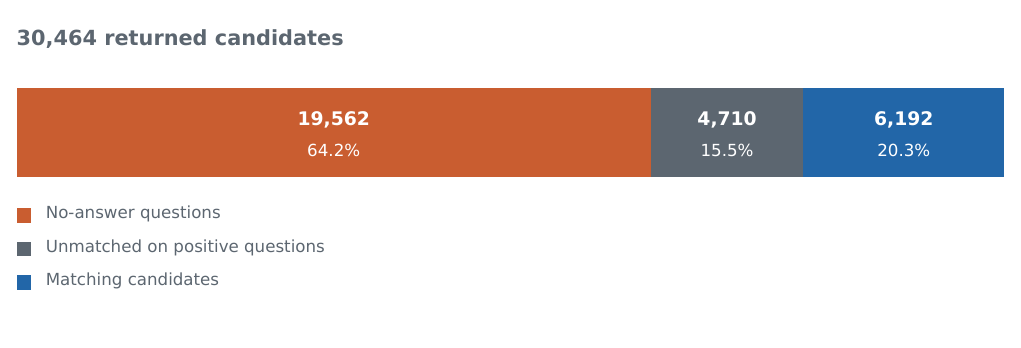}
\caption{Composition of the corrected output. The 6,192 matching candidate strings collectively recover 2,172 references. A candidate string and a recovered reference are different counting units.}\label{fig:composition}
\end{figure}

Of 2,938 questions with no annotated answer, 1,335 receive at least one candidate. Their \NoAnswerRate{} false-positive rate has a 95\% contract-bootstrap interval of 43.3--47.9\%. These questions contribute 19,562 candidate strings: \NoAnswerShare{} of all output and \NoAnswerFPShare{} of unmatched candidates. Figure~\ref{fig:composition} separates this output from unmatched candidates on positive questions and candidates that match a reference.

Six categories answer every negative test question. Examples are \emph{Competitive Restriction Exception} (86/86), \emph{Non-Disparagement} (95/95), and \emph{Minimum Commitment} (70/70), each with an observed no-answer FPR of 100\%. All three use the minimum-cutoff fallback. These are sample rates, not claims of certain future error. Minimum Commitment also meets both 30-contract support minima, showing that passing the support screen does not imply satisfactory abstention. Appendix~\ref{app:categories} reports all categories.

Erroneous answering is a substantial source of output at this operating point, but its concentration does not establish the effectiveness of an intervention. Negative questions account for 70.3\% of the test questions, categories use different cutoffs, and the minimum-cutoff fallback can favor broad output. The results motivate a separately designed abstention experiment; they do not show that it would outperform improved span ranking, overlap consolidation, or a different threshold policy.

\subsection{Category support and its sensitivity}
\begin{figure}[p]\centering
\includegraphics[width=.91\linewidth]{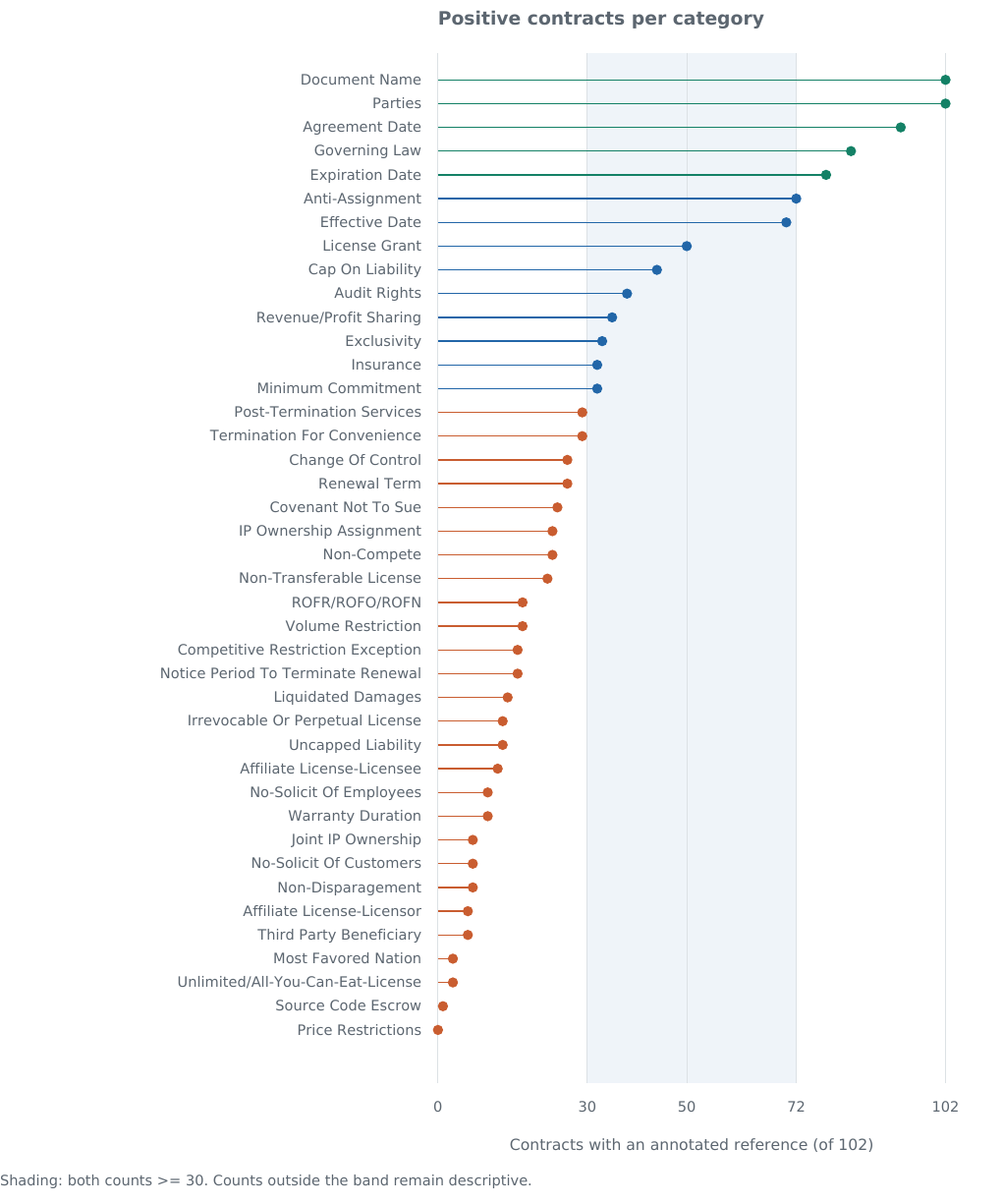}
\caption{Positive-contract support for each category. Nine categories lie in the shaded window for the chosen minimum of 30 positives and 30 negatives. Outside that window, estimates with nonzero denominators remain descriptive. Document Name and Parties have zero negatives; Price Restrictions has zero positives.}\label{fig:support}
\end{figure}

At the prespecified minimum of 30 contracts per arm, nine of 41 categories pass both arms, five have adequate positive support only, and 27 have adequate negative support only. Because each category is queried once per contract, $n_- = 102-n_+$, and the two-arm screen becomes $30\leq n_+\leq72$ (Figure~\ref{fig:support}). The passing count is 18 with a minimum of 20 and two with a minimum of 40. It is therefore a property of the selected support requirement and sample, not an intrinsic count of categories whose performance is measurable.

\emph{Document Name} and \emph{Parties} appear in every test contract. Their no-answer false-positive rates are undefined, although their span-based metrics remain computable. \emph{Agreement Date}, \emph{Governing Law}, and \emph{Expiration Date} have 9, 19, and 24 negative contracts. Their no-answer rates are estimable but below the selected support minimum. For example, Governing Law returns no candidate on any of its 19 negative questions; a 95\% Wilson upper limit is approximately 16.8\%, illustrating why zero observed errors do not establish a negligible false-positive rate.

Of the 541 original registry rows, 22 pass their declared metric-specific screen: 16 for both arms, five for recall only, and one for no-answer analysis only. The other 519 original rows and all eight supplemental S3N rows fail their declared screen. The 404 registered intersection rows are all below their support requirement. These rows overlap and are not independent replications. Their high failure fraction partly reflects the number and granularity of planned slices, including the fragmented title-derived type field.

\clearpage
\subsection{Training-frequency and length slices}
\begin{figure}[htbp]\centering
\includegraphics[width=.94\linewidth]{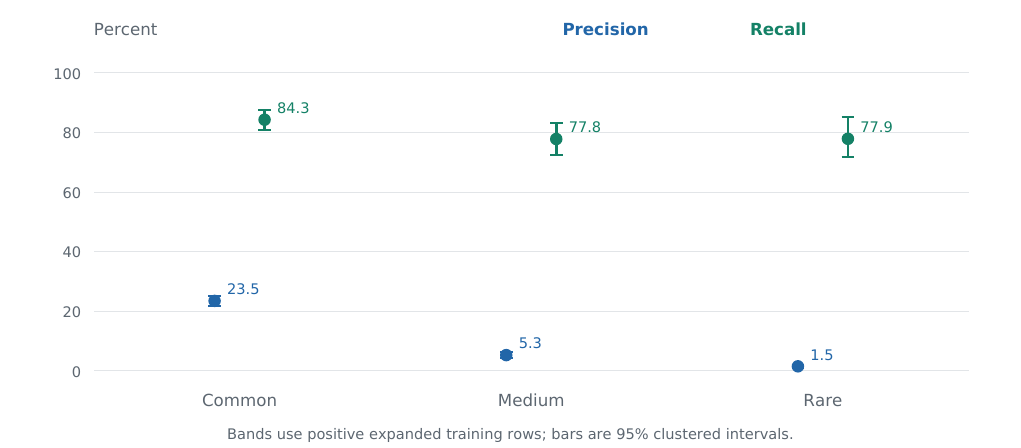}
\caption{Precision and recall by training-frequency band. Error bars are marginal 95\% contract-bootstrap intervals; they do not isolate a causal effect of frequency.}\label{fig:frequency}
\end{figure}

Table~\ref{tab:frequency} reports the original training-frequency bands under corrected scoring. Precision is lower in the lower-frequency bands, while observed recall is 84.3\% for common, 77.8\% for medium, and 77.9\% for rare. Figure~\ref{fig:frequency} includes clustered uncertainty. We do not infer that frequency causes the precision differences or excludes an effect of category difficulty: training support, reference multiplicity, semantics, and thresholds vary together.

\begin{table}[htbp]\centering\small
\caption{Frozen S5 bands, correctly identified as positive expanded training-row frequency. P denotes CUAD precision and R reference recall.}\label{tab:frequency}
\begin{tabular}{lrrrrr}
\toprule
Training frequency & Questions & References & Returned & P (\%) & R (\%) \\
\midrule
Common & 1,326 & 1,777 & 8,848 & 23.5 & 84.3 \\
Medium & 1,428 & 667 & 10,995 & 5.3 & 77.8 \\
Rare & 1,428 & 199 & 10,621 & 1.5 & 77.9 \\
\bottomrule
\end{tabular}

\end{table}

The rare band returns 10,621 candidate strings for 199 references, compared with 8,848 strings for 1,777 references in the common band. The bands contain 14, 14, and 13 categories for rare, medium, and common respectively; they therefore differ slightly in question exposure as well as in reference counts. Per-category counts in Appendix~\ref{app:categories} permit inspection of this heterogeneity.

The corrected longest-reference-length slice also does not support a null effect: precision across its four groups is 58.1\%, 40.2\%, 24.7\%, and 25.7\%. These groups contain only positive questions and are defined by each question's longest reference, with cut points of 6, 31, and 57 whitespace tokens. The Parties substring rule contributes to the scoring correction for this slice. We report the association descriptively, without an equivalence or causal claim. Document-length groups contain 29, 35, 17, and 21 contracts; only the second group meets the original support screen. Appendix~\ref{app:other} provides both sets of summaries.

\section{Discussion and limitations}
The study demonstrates why an evaluation report should distinguish \emph{what is counted} from \emph{how well a system performs}. A recovered reference, a matching candidate, and a correctly declined question each answer a different question about the review workflow. The difference between \MainPrecision{} CUAD precision and \CandidateMatch{} candidate match fraction is not a contradiction. It follows from the matching convention and the presence of overlapping candidate variants.

The positive/negative support inventory is similarly a diagnostic description, not a release decision. Specifying an acceptable no-answer error rate or recall shortfall would require a decision context and an uncertainty criterion; selecting 30 contracts by convention does not supply either. Categories with no negative support also raise a task-design question: where a field is structurally present in the intended population, extraction quality may be the relevant target, whereas no-answer behavior would need deliberately different inputs to be evaluated.

Several limits restrict generalization. This is one older extractive checkpoint on a public benchmark of commercial contracts. It does not characterize current language models or other review systems. The decoder retains at most 20 locations and thresholds can return overlapping variants. Candidate counts are consequently properties of the full configuration, and the attainable recall of the saved candidate pool is not a bound on the architecture under other decoding choices.

The thresholds are selected on a small, expanded training-split sample whose independence from checkpoint training is unverified. Many categories have few selection references, and the 12 minimum-cutoff fallbacks can dominate output volume. Neither training overlap nor selection uncertainty is removed by resampling test contracts. Earlier test access and retrospective corrections further preclude a claim of independent prospective validation.

CUAD annotations are the reference standard for this experiment, not an exhaustive determination of legal meaning. A question with no annotated span is called a no-answer question here; annotation omissions or ambiguity could change its interpretation. No qualified legal reviewer adjudicated the errors, and no human study measured usefulness or review time. The original automated failure taxonomy was screening only; the corrected paper does not assign rates to unadjudicated contextual modes such as negation, cross-reference, or truncation errors.

All categories were designated human-review-only in the original plan. Zero autonomous coverage and universal human review were therefore design constraints, not empirical evidence that a particular release policy succeeded. The present results do not establish compliance, safety, production readiness, or portability to another document domain. A useful follow-up would specify a concrete no-answer intervention and an operating objective in advance, establish data independence, and evaluate on new evidence.

\clearpage
\section{Reproducibility and conclusion}
The ancillary directory \texttt{anc/} includes the recorded inputs needed to recompute the corrected metrics, original frozen code and manifests, the correction plan, a second scoring implementation, generated metrics, and figure-generation code. To reduce download size, source data and predictions use a lossless structural encoding: contract text is stored once and repeated passages are represented by exact character spans. The supplied restoration script recovers all four original input files byte for byte and verifies their original hashes before rescoring. The correction runs with Python~3.12.14 and NumPy~2.3.5; figures use ReportLab~4.4.9. Historical runtime versions are retained as recorded in the original manifest and are not requirements for rescoring.

Software verification compares all 4,182 questions under three analyses (12,546 count comparisons), checks 8,364 question-level precision/recall results against the upstream CUAD functions for the two corrected-scoring analyses, verifies all 40 selection categories with references, and independently reproduces four headline bootstrap intervals. The no-reference category's abstaining policy is also checked. These are software cross-checks performed in this revision, not external peer review. The upstream source revision and hash are recorded for repeat verification. Data excerpts and annotations retain CUAD attribution and its CC BY 4.0 terms.

The corrected case study identifies substantial output on no-answer questions and uneven category support while showing how scorer conventions affect the reported results. Its practical contribution is a traceable account of those measurements and their boundaries. It supports further evaluation of abstention and operating policies; it does not supply a general release criterion.

\paragraph{Author's views.}
The views expressed in this article are those of the author alone and do not necessarily reflect the views of any current or former employer or affiliated organization.

\clearpage
\appendix
\section{Correction record and provenance}\label{app:correction}
Table~\ref{tab:corrections} separates the effect of corrected matching from threshold reselection. Both corrected analyses use the same saved candidates. Only Document Name changes its cutoff, adding 133 returned strings and recovering six additional references relative to the scoring-only correction. The original files remain unchanged; the main text uses the final row throughout.

\begin{table}[htbp]\centering\small
\caption{Original and corrected analyses. TP counts recovered references; FP counts unmatched candidates. P and R are CUAD precision and reference recall.}\label{tab:corrections}
\begin{tabular}{lrrrrrr}
\toprule
Analysis & TP & FP & FN & Returned & P (\%) & R (\%) \\
\midrule
Original & 2,053 & 25,275 & 590 & 30,331 & 7.51 & 77.68 \\
Matching corrected & 2,166 & 24,215 & 477 & 30,331 & 8.21 & 81.95 \\
Matching + thresholds & 2,172 & 24,272 & 471 & 30,464 & 8.21 & 82.18 \\
\bottomrule
\end{tabular}

\end{table}

The original implementation collapsed whitespace and lacked the Parties substring exception; its draft description additionally claimed article removal that the code did not perform. Literal-space matching and the substring allowance account for the change between the first two rows. An earlier draft also mentioned an unexplained independent count of 2,075 recovered references. That preliminary number is not used as evidence here; the exact original counts and current corrected counts have been independently reproduced, replacing the unresolved approximate-reproduction claim.

The original five-category ranking assertion and the causal interpretation of training frequency are withdrawn. The former did not match the original results; the latter conflated expanded annotation frequency with contract prevalence. The claimed span-length null does not follow from either the original or corrected analysis. Workload statements now specify candidate and reference denominators separately. The corrected validation-row and feature-window totals are reported in Section~\ref{sec:design}.

The recorded acquisition timestamp is 20 August 2026 at 23:21:33 UTC. The pre-execution addendum predates the recorded final freeze, including its two-arm support rule and permission to enumerate test-label support. The surviving audit documents do not fully resolve the earlier calibration access; this uncertainty is retained rather than converted into a never-accessed-holdout claim. The original protocol and audits are included as historical documents, with a correction note identifying descriptions superseded by this paper.

\begin{table}[htbp]\centering\small
\caption{Selected original artifact hashes. Full hashes and all corrected output hashes are in the ancillary manifest.}
\begin{tabular}{ll}\toprule Artifact & SHA-256 prefix\\\midrule
Original dataset manifest & \texttt{8306be059da8}\\
Original threshold record & \texttt{d0287c3b7da0}\\
Original run manifest & \texttt{397c0fdccfa5}\\
Original test predictions & \texttt{e564a78fb74c}\\
Original result seal & \texttt{016bdbf0eff6}\\\bottomrule
\end{tabular}
\end{table}

\clearpage
\section{Additional descriptive summaries}\label{app:other}
\begin{table}[htbp]\centering\small
\caption{Length slices under corrected scoring. Document cut points are 2,512, 5,358, and 10,497 whitespace tokens. Reference cut points are 6, 31, and 57, using the longest reference on each positive question. Q1 is the shortest group.}
\begin{tabular}{lrrrr}
\toprule
Length group & Contracts & Questions & P (\%) & R (\%) \\
\midrule
Document Q1 & 29 & 1,189 & 4.5 & 84.2 \\
Document Q2 & 35 & 1,435 & 7.8 & 89.3 \\
Document Q3 & 17 & 697 & 8.5 & 82.9 \\
Document Q4 & 21 & 861 & 12.2 & 76.1 \\
Reference Q1 & 102 & 286 & 58.1 & 88.0 \\
Reference Q2 & 91 & 260 & 40.2 & 92.0 \\
Reference Q3 & 81 & 292 & 24.7 & 84.0 \\
Reference Q4 & 76 & 406 & 25.7 & 74.1 \\
\bottomrule
\end{tabular}

\end{table}

\begin{table}[htbp]\centering\small
\caption{Descriptive ranking measures using null-relative candidate scores. Conditional means exclude undefined categories and show their denominators. These are not direct replications of the published probability-sweep baseline.}\label{tab:ranking}
\begin{tabular}{lrrr}
\toprule
Measure & Pooled & Conditional category mean & Defined categories \\
\midrule
AUPR & 53.7\% & 53.3\% & 40/41 \\
Precision at 80\% recall & 30.3\% & 35.2\% & 33/41 \\
Precision at 90\% recall & -- & 27.2\% & 19/41 \\
\bottomrule
\end{tabular}

\end{table}

The ranking calculation evaluates every distinct saved candidate score, applies a monotone precision envelope, and integrates by the trapezoidal rule from recall zero. It adds no area beyond the maximum attainable recall. The saved candidate pool reaches 88.3\% pooled recall, so pooled precision at 90\% recall is undefined. The official CUAD baseline evaluator instead sweeps candidate probabilities over a fixed grid; matching compatibility does not make these ranking procedures identical. Macro precision at a target recall is conditional on attaining that target, not a performance average over all 41 categories. Categories with no references and those failing to attain a target are separately represented in the machine-readable report.

At the corrected operating point, means over categories with defined metrics are 23.2\% precision (40 categories), 78.1\% recall (40), and 29.1\% F1 (41). Precision is undefined for Source Code Escrow because it returns no candidates; recall is undefined for Price Restrictions because the test set has no references. The direct count formula for F1 is defined as zero in both cases. These means weight categories equally and answer a different question from pooled performance.

\clearpage
\section{All category results}\label{app:categories}
\begin{table}[!htbp]\centering\footnotesize
\caption{Corrected test results for all 41 categories. $n_+$ and $n_-$ count positive and negative contracts; P and R are CUAD precision and reference recall. FPR is the question-level no-answer rate. Dashes indicate undefined denominators. Full intervals and counts are in the ancillary report; small-support rows remain descriptive.}\label{tab:categories}
\begin{tabular}{lrrrrr}
\toprule
Category & $n_+$ & $n_-$ & P (\%) & R (\%) & FPR (\%) \\
\midrule
Affiliate License-Licensee & 12 & 90 & 1.3 & 92.6 & 98.9 \\
Affiliate License-Licensor & 6 & 96 & 0.7 & 70.0 & 100.0 \\
Agreement Date & 93 & 9 & 65.5 & 83.9 & 33.3 \\
Anti-Assignment & 72 & 30 & 41.6 & 84.7 & 20.0 \\
Audit Rights & 38 & 64 & 38.8 & 82.9 & 7.8 \\
Cap On Liability & 44 & 58 & 31.4 & 78.8 & 32.8 \\
Change Of Control & 26 & 76 & 2.5 & 72.6 & 98.7 \\
Competitive Restriction Exception & 16 & 86 & 1.0 & 70.4 & 100.0 \\
Covenant Not To Sue & 24 & 78 & 56.0 & 63.6 & 3.8 \\
Document Name & 102 & 0 & 31.5 & 82.4 & -- \\
Effective Date & 70 & 32 & 26.2 & 75.0 & 75.0 \\
Exclusivity & 33 & 69 & 18.2 & 83.3 & 27.5 \\
Expiration Date & 78 & 24 & 48.3 & 88.0 & 20.8 \\
Governing Law & 83 & 19 & 96.1 & 82.2 & 0.0 \\
Insurance & 32 & 70 & 55.9 & 76.9 & 2.9 \\
IP Ownership Assignment & 23 & 79 & 22.8 & 47.5 & 24.1 \\
Irrevocable Or Perpetual License & 13 & 89 & 17.4 & 91.3 & 24.7 \\
Joint IP Ownership & 7 & 95 & 16.9 & 92.9 & 18.9 \\
License Grant & 50 & 52 & 37.4 & 77.8 & 11.5 \\
Liquidated Damages & 14 & 88 & 12.2 & 87.0 & 31.8 \\
Minimum Commitment & 32 & 70 & 3.9 & 80.7 & 100.0 \\
Most Favored Nation & 3 & 99 & 1.5 & 100.0 & 27.3 \\
No-Solicit Of Customers & 7 & 95 & 9.2 & 80.0 & 17.9 \\
No-Solicit Of Employees & 10 & 92 & 33.3 & 94.4 & 2.2 \\
Non-Compete & 23 & 79 & 35.2 & 74.6 & 20.3 \\
Non-Disparagement & 7 & 95 & 0.6 & 91.7 & 100.0 \\
Non-Transferable License & 22 & 80 & 10.4 & 81.4 & 38.8 \\
Notice Period To Terminate Renewal & 16 & 86 & 34.1 & 77.8 & 12.8 \\
Parties & 102 & 0 & 82.3 & 93.9 & -- \\
Post-Termination Services & 29 & 73 & 2.9 & 65.9 & 98.6 \\
Price Restrictions & 0 & 102 & 0.0 & -- & 2.0 \\
Renewal Term & 26 & 76 & 17.1 & 96.8 & 21.1 \\
Revenue/Profit Sharing & 35 & 67 & 14.1 & 79.3 & 28.4 \\
ROFR/ROFO/ROFN & 17 & 85 & 3.0 & 82.4 & 98.8 \\
Source Code Escrow & 1 & 101 & -- & 0.0 & 0.0 \\
Termination For Convenience & 29 & 73 & 8.6 & 90.2 & 57.5 \\
Third Party Beneficiary & 6 & 96 & 40.0 & 18.2 & 2.1 \\
Uncapped Liability & 13 & 89 & 8.3 & 81.2 & 32.6 \\
Unlimited/All-You-Can-Eat-License & 3 & 99 & 0.3 & 83.3 & 99.0 \\
Volume Restriction & 17 & 85 & 1.7 & 91.4 & 100.0 \\
Warranty Duration & 10 & 92 & 0.8 & 78.9 & 100.0 \\
\bottomrule
\end{tabular}

\end{table}
\end{document}